\newif\ifsingle

\documentclass[a4paper,conference]{IEEEtran}
\IEEEoverridecommandlockouts

\usepackage[top=1.9cm, bottom=4.3cm,left=1.3cm,right=1.3cm]{geometry}
\def\BibTeX{{\rm B\kern-.05em{\sc i\kern-.025em b}\kern-.08em T\kern-.1667em\lower.7ex\hbox{E}\kern-.125emX}}
\usepackage[noadjust]{cite}
\usepackage[T1]{fontenc}
\usepackage[latin9]{inputenc}
\usepackage{color}
\usepackage{amsmath}
\usepackage{amssymb}
\usepackage{graphicx}
\usepackage{textcomp}
\usepackage{xcolor}
\usepackage{balance}
\usepackage[bookmarks,hidelinks]{hyperref}

\makeatletter

\providecommand{\tabularnewline}{\\}

\ifsingle
\newcommand{\figWidth}{0.65\columnwidth} 
\else
\newcommand{\figWidth}{1\columnwidth} 
\fi 

\definecolor{NewColor}{rgb}{0.2,0,0.5}

\makeatother

\begin{document}
\title{Transformer-Based Deep Learning Detector for Dual-Mode Index Modulation 3D-OFDM}

\author{Toan Gian, Tien-Hoa Nguyen, Trung Tan Nguyen, Van-Cuong Pham, and Thien Van Luong

\thanks{Toan Gian and Tien-Hoa Nguyen are with the School of Electrical and Electronics Engineering, Hanoi University of Science and Technology, Hanoi, Vietnam (e-mail: Toandinh7176@gmail.com, \{hoa.nguyentien \}@hust.edu.vn.)}

\thanks{Trung Tan Nguyen is with the Faculty of Radio-Electronics, Le Quy Don Technical University, Ha Noi 11355, Vietnam (e-mail: trungtannguyen@mta.edu.vn).}

\thanks{Van-Cuong Pham, Thien Van Luong are with the Faculty of Computer Science, Phenikaa University, Hanoi 12116, Vietnam (e-mail: \{cuong.phamvan, thien.luongvan\}@phenikaa-uni.edu.vn). }




\vspace{-0.5cm}
 }

\maketitle
\vspace{-1cm}

\begin{abstract}
In this paper, we propose a deep learning-based signal detector called TransD3D-IM, which employs the Transformer framework for signal detection in the Dual-mode index modulation-aided three-dimensional (3D) orthogonal frequency division multiplexing (DM-IM-3D-OFDM) system. In this system, the data bits are conveyed using dual-mode 3D constellation symbols and active subcarrier indices. As a result, this method exhibits significantly higher transmission reliability than current IM-based models with traditional maximum likelihood (ML) detection. Nevertheless, the ML detector suffers from high computational complexity, particularly when the parameters of the system are large. Even the complexity of the Log-Likelihood Ratio algorithm, known as a low-complexity detector for signal detection in the DM-IM-3D-OFDM system, is also not impressive enough. To overcome this limitation, our proposal applies a deep neural network at the receiver, utilizing the Transformer framework for signal detection of DM-IM-3D-OFDM system in Rayleigh fading channel. Simulation results demonstrate that our detector attains to approach performance compared to the model-based receiver. Furthermore, TransD3D-IM exhibits more robustness than the existing deep learning-based detector while considerably reducing runtime complexity in comparison with the benchmarks.
\end{abstract}



\section{Introduction}
Combining orthogonal frequency division multiplexing (OFDM) with index modulation (IM) methods, namely OFDM-IM \cite{barsar2013ofdmim}, \cite{ThienTVT2017} has gained dominant attention as a potential approach to substitute the traditional OFDM technique. In OFDM-IM, a subset of subcarriers is activated to transmit data through both subcarriers and their corresponding indices. As a result, the OFDM-IM approach outperforms for both reliability and energy efficient compared to traditional methods thanks to using indices of active subcarriers to convey data. Additionally, adjusting the number of activated subcarriers promises to offer an flexible trade-off between the error performance and spectral efficiency (SE).

In recent years, various advanced OFDM-based systems have been introduced \cite{Ko2014tightupper, Pout2017, Basar2016mimoIM, thienWCL2018}, aiming at enhancing either the error performance or SE. Additionally, the dual-mode OFDM (DM-OFDM) \cite{Mao2017} and three-dimensional OFDM (3D-OFDM) \cite{kang20083dofdm} are also known as the efficiency approach in reducing the bit error rate (BER) of the traditional OFDM. In particular, DM-OFDM utilizes multiple constellations to convey data using subcarriers that are not activated. In comparison with OFDM-IM, DM-OFDM has shown superior in higher achievable SE. A generalized variant of DM-OFDM which greatly improves the SE of the traditional version in \cite{wen2017mmIM}. In the process of 3D-OFDM, data bits are converted to the 3D constellation symbols and the obtained set of symbols is divided into subcarriers, which are employed the 2D inverse fast Fourier transform for modulation. By employing OFDM with a 3D mapper instead of a 2D one, the minimum Euclidean distance among different constellations notably increases. These enhancement results in error performance thereby reduce the likelihood of transmission errors. In \cite{chen2010clo} and \cite{huang2018papr}, authors presented the analysis for symbol error probability (SEP) and the solution to decrease the peak-to-average power ratio of 3D-OFDM. Nevertheless, its SE is only one-third compared to the classical OFDM when considering the constellation order \cite{CHEN2020impro}. Motivated by the IM idea, a novel scheme combined by 3D-OFDM \cite{kang20083dofdm} and DM-OFDM \cite{Mao2017}, namely dual-mode IM-aided 3D-OFDM (DM-IM-3D-OFDM), was introduced in \cite{Wang2021dm3d} to enhance SE of 3D-OFDM. Within this model, data bits are conveyed via indices of activated subcarriers and the 3D mapper constellations based on the floor of the Poincare sphere \cite{Chen2011}. In particular, subcarriers are divided into subblocks and subcarriers in each subblock are split into two groups, then mapped by dual-mode 3D constellations. In comparison with 3D-OFDM, DM-IM-3D-OFDM demonstrates superior to transmission reliability at large SNRs when employing the maximum likelihood (ML) detection under the fading channels. However, its computational complexity is high, especially as parameters of model are large. Even the complexity of the Log-Likelihood Ratio (LLR) algorithm, known as a low-complexity detector for signal detection in the DM-IM-3D-OFDM system, is also not impressive enough. Consequently, we propose a new detector to solve this fundamental issue.

Recently, Deep Learning (DL) \cite{SCHMIDHUBER201585} has been broadly applied to the physical layer of wireless communication systems \cite{Thien2019DLIM, Luong2020engery, Luong2020MC-AE, Luong2022optical, Luong2022SIC, Toan2022APSIPA}. Furthermore, the DL-based detector successfully employs the Transformer framework \cite{attention} for signal detection in the OFDM-IM system, especially when the model parameters are large \cite{TransIM}. To the best of our knowledge, the application of DL Transformer architecture to the DM-IM-3D-OFDM has been overlooked in the literature. In this contribution, our proposal focuses on a novel Transformer-based DL detector called TransD3D-IM for signal detection of the DM-IM-3D-OFDM system. In particular, we build a new deep neural network (DNN) model of TransD3D-IM based on the Transformer framework. In our proposal, the Transformer block plays a crucial role as it calculates global dependencies and features with the 3D signal, which allows to obtain BER performance improvement. Aiming of minimizing BER, TransD3D-IM is trained offline by employing the simulation dataset. After that, it plays the role of an online detector to quickly estimate the conveyed bits. Our simulation results pointed out that TransD3D-IM can achieve an approach BER performance and much lower runtime complexity than the previous detectors.

The remaining parts of the paper are organized as follows. Section~\ref{sec:System-Model} describes the system model of DM-IM-3D-OFDM and conventional ML and LLR detectors. Section~\ref{sec:DeepNet} presents the structure and model training of our proposed TransD3D-IM detector. Subsequently, Section~\ref{sec:SIMULATION RESULTS} shows the results achieved by our transformer-based detector and compares it with conventional detectors. Lastly, Section~\ref{sec:Conclusions} is conclusion.


\section{system model\label{sec:System-Model}}

\begin{figure}[tb]
\begin{centering}
\includegraphics[width=\figWidth]{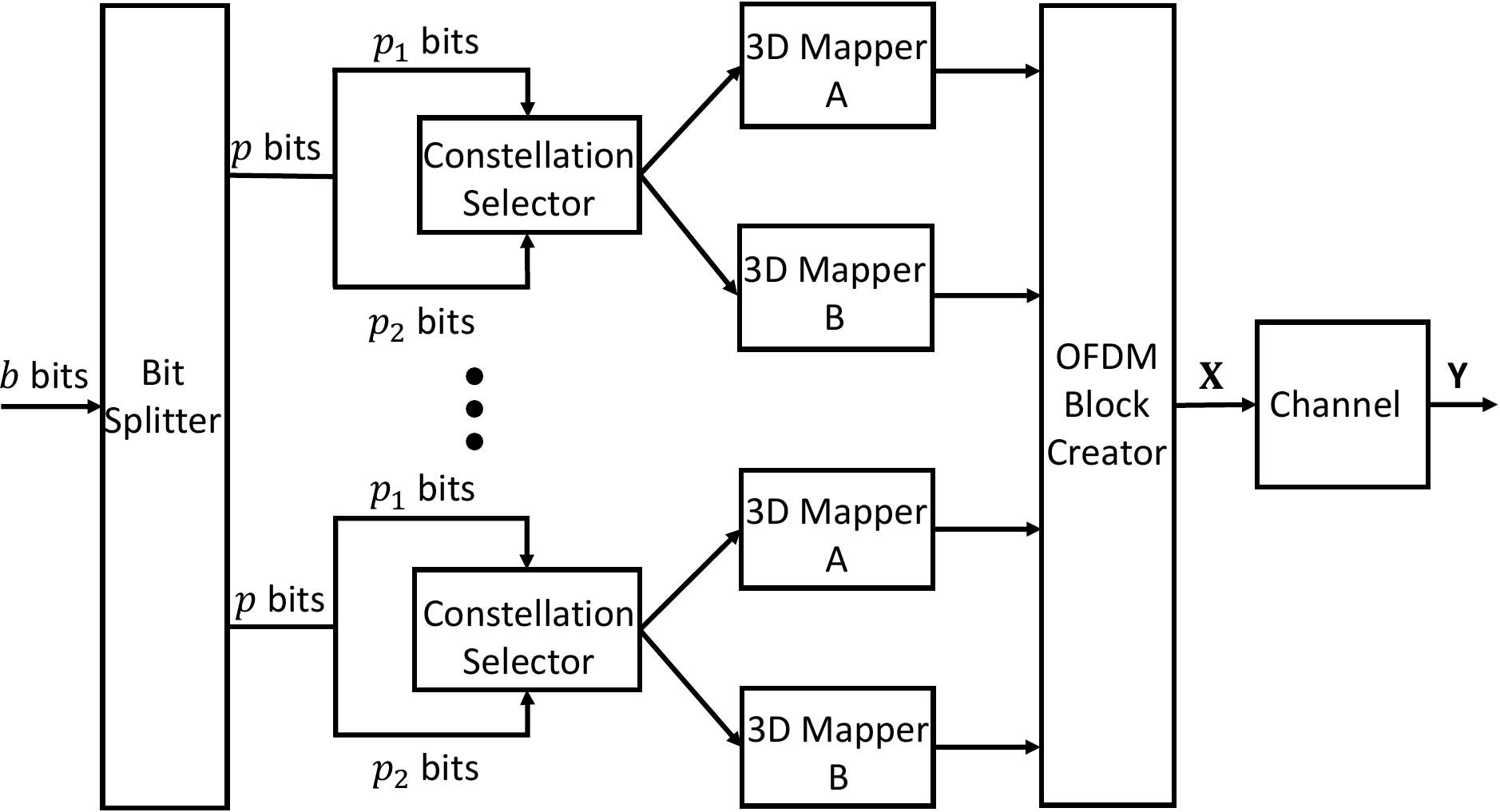} 
\par\end{centering}
\caption{Block diagram of the DM-IM-3D-OFDM system.\label{fig:figure1}}
\end{figure}

\vspace{-0.1cm}
The diagram in Fig.~\ref{fig:figure1} illustrates the structure of a DM-IM-3D-OFDM simplified system, where data are conveyed by mapping 3D constellations and subcarrier indices. It is assumed that the message consisting of $b$ bits are split into $m$ subblocks of $n$ subcarriers and $p$ bits, i.e. $m = b/p=N/n$, where $N$ is the number of subcarriers per subblock. Herein, the $p$ bits are split into two components based on the dual 3D constellation-IM scheme. The first component, referred to as the index bit, consists of $p_{1}$ bits. The index bit serves as a selector to divide the subcarrier indices into two subsets. The second component, known as the information bits, consists of $p_{2}$ bits and is mapped to two distinct constellations: type A and type B, denoted by $\mathbf{\mathcal{S}}_{A}$ and $\mathbf{\mathcal{S}}_{B}$, respectively. Specifically, symbol sets of $\mathbf{\mathcal{S}}_{A}$ and $\mathbf{\mathcal{S}}_{B}$ are represented by $\mathbf{\mathcal{M}}_{A}$ and $\mathbf{\mathcal{M}}_{B}$ that associated with the size of $s_{A}$ and $s_{B}$, respectively, note that $\mathbf{\mathcal{M}}_{A} \cap \mathbf{\mathcal{M}}_{B} = \varnothing$. In the frequency domain, each element of $\mathbf{\mathcal{M}}_{A}$ can be expressed by $\mathbf{\mathcal{M}}_{A}^{j}=(x,y,z)^{T}$, i.e., $j$-th symbol from the symbol set $\mathbf{\mathcal{M}}_{A}$ where $1 \leq j \leq s_A$ and $x,y,z$ are coefficients of base vectors in $x-$, $y-$, and $z-$axis. In addition, $(.)^{T}$ presents the transposition. Value of $k$ stands for the number of subcarriers modulated by $\mathbf{\mathcal{S}}_{A}$, thus the number of subcarriers mapped to $\mathbf{\mathcal{S}}_{B}$ is $n-k$. 
Accordingly, we can determine the value of $p$ as follows.

\begin{equation}
    p = p_{1} + p_{2} = \lfloor \log_2C_n^k \rfloor + \log_2((s_A)^k) + \log_2((s_B)^{n-k}),
\end{equation}
where $C_n^k$ is the binomial coefficient and $\lfloor.\rfloor$ denotes the integer floor operator. Additionally, the index selectors of the two 3D mapper constellations have an interdependent relationship. Specifically, once the index selector of $\mathbf{\mathcal{S}}_{A}$ is determined, the corresponding index selector of $\mathbf{\mathcal{S}}_{B}$ is also known. Therefore, it is sufficient to define the index selector of $\mathbf{\mathcal{S}}_{A}$ in order to describe the principle of the index selector. For instance, it can be selected that $n = 4$, $k = 2$, and $s_A=s_B=2$ for each OFDM subblock. The IM procedure is presented in Table~\ref{tab:table1}.

\begin{table}[!ht]
\centering
\caption{The Look-up Table for DM-IM-3D-OFDM System\label{tab:table1}}
\begin{tabular}{|c|c|c|}
\hline
IM Bits & Indices for $\mathbf{\mathcal{S}}_{A}$ & Subblocks \\ \hline
$[0,0]$           & $[1,2]$    & $[\mathbf{\mathcal{S}}_{A}^{1},\mathbf{\mathcal{S}}_{A}^{2},\mathbf{\mathcal{S}}_{B}^{1},\mathbf{\mathcal{S}}_{B}^{2}] $                                              \\ \hline
$[0,1]$           & $[2,3]$    & $[\mathbf{\mathcal{S}}_{B}^{1},\mathbf{\mathcal{S}}_{A}^{1},\mathbf{\mathcal{S}}_{A}^{2},\mathbf{\mathcal{S}}_{B}^{2}]$                                               \\ \hline
$[1,0]$           & $[3,4]$    & $[\mathbf{\mathcal{S}}_{B}^{1},\mathbf{\mathcal{S}}_{B}^{2},\mathbf{\mathcal{S}}_{A}^{1},\mathbf{\mathcal{S}}_{A}^{2}]$                                               \\ \hline
$[1,1]$           & $[1,4]$    & $[\mathbf{\mathcal{S}}_{A}^{1},\mathbf{\mathcal{S}}_{B}^{1},\mathbf{\mathcal{S}}_{B}^{2},\mathbf{\mathcal{S}}_{A}^{2}]$                                               \\ \hline
\end{tabular}
\end{table}

In the frequency domain, the transmitted signal $\mathbf{X} \in \mathbb{C}^{3 \times n} $ in a block is constructed by $\mathbf{X}=[\mathbf{X}(1),\mathbf{X}(2),\ldots,\mathbf{X}(n)]$, where $\mathbf{X}(i) \in \left\{\mathbf{\mathcal{S}}_{A}, \mathbf{\mathcal{S}}_{B}\right\}$ and $1 \leq i \leq n$. Particularly, the output of two mappers can be represented by $\mathbf{\mathcal{S}}_{A}=[\mathbf{\mathcal{S}}_{A}^{1},\ldots,\mathbf{\mathcal{S}}_{A}^{k}]$, where $\mathbf{\mathcal{S}}_{A}^{i}=\mathbf{\mathcal{M}}_{A}^{j}$ $(1 \leq i \leq n, 1 \leq j \leq s_A)$ and $\mathbf{\mathcal{S}}_{B}=[\mathbf{\mathcal{S}}_{B}^{1},\ldots,\mathbf{\mathcal{S}}_{B}^{n-k}]$, where $\mathbf{\mathcal{S}}_{B}^{i}=\mathbf{\mathcal{M}}_{B}^{j}$ $(1 \leq i \leq n-k, 1 \leq j \leq s_B)$. Noting that the process of designing 3D constellations to built-up the signal matrix $\mathbf{X}$ is presented in detail in \cite{Wang2021dm3d}, the index selector may be obtained by the look-up table as shown in Table~\ref{tab:table1}. 

At the receiver, the obtained signal matrix $\mathbf{Y} \in \mathbb{C}^{3 \times n}$ in the frequency domain can be represented as follows.
\begin{equation}
    \mathbf{Y} = \mathbf{H}\mathbf{X} + \mathbf{G},
\end{equation}
where $\mathbf{H} = [\mathbf{H}_1;\mathbf{H}_2;\mathbf{H}_3]^T \in \mathbb{C}^{3\times n},\mathbf{H}_{i}\in \mathbb{C}^{1 \times n}$ and $\mathbf{G}=[\mathbf{G}_1,\mathbf{G}_2,\mathbf{G}_3]^T \in \mathbb{C}^{3 \times n}, \mathbf{G}_{i}\in \mathbb{C}^{1 \times n}$ $(1\leq i \leq 3)$, represent the fading channel coefficients and the additive white Gaussian noise (AWGN), respectively. Each entry of $\mathbf{H}$ and $\mathbf{G}$ are identically and independently distributed as a complex-valued random variable, with $\mathbf{H}_{i}(\alpha)\sim\mathcal{C}\mathcal{N}(0,1)$ and $\mathbf{G}_{i}(\alpha)\sim\mathcal{C}\mathcal{N}(0, N_{0})$ $(1 \leq \alpha \leq n)$, respectively, where $N_{0}$ is the noise variance.



At the received side, the ML detector is employed and the estimated signal denoted by $\mathbf{\hat{X}}$ is calculated by
\begin{equation}
\
\hat{\mathbf{X}}=\arg\,\underset{\mathbf{X}\in\delta}{\min}\parallel\mathbf{Y}-\mathbf{H}\mathbf{X}\parallel^{2},
\, \label{eq:ML_equation}
\end{equation}
where $\parallel.\parallel^{2}$ is the squared Euclidean norm and $\delta$ stands for the set of all possible candidate delivery vectors modulated with $\mathbf{\mathcal{S}}_{A}$ and $\mathbf{\mathcal{S}}_{B}$.
While ML can yield good performance, it becomes impractical as the values of $p_{1}$ and $k$ increase due to its exponential growth.
Moreover, $\mathbf{\hat{X}}$ can be recovered through the LLR detector by considering that the subcarriers are modulated by either $\mathbf{\mathcal{S}}_{A}$ or $\mathbf{\mathcal{S}}_{B}$, the LLR algorithm provides the ratio of the posterior probabilities, which can be expressed by

\begin{equation}
\
\delta_{i}=\ln\left(\frac{\mathop{\sum_{j=1}^{s_{A}}P(\mathbf{X}(i)=\mathbf{\mathcal{M}}_{A}^{j})\mid\mathbb{\mathbf{Y}}(i)}}{\sum_{j=1}^{s_{B}}P(\mathbf{X}(i)=\mathbf{\mathcal{M}}_{B}^{j})\mid\mathbf{Y}(i)}\right)
\,, \label{eq:LLR}
\end{equation}
where $1\leq i\leq n$, $\mathbf{\mathcal{M}}_{A}^{j} \in \mathbf{\mathcal{M}}_{A}$ and $\mathbf{\mathcal{M}}_{B}^{j} \in \mathbf{\mathcal{M}}_{B}$. As observed in \eqref{eq:LLR}, the value of $\delta_{i}$ determines whether the $i$-th subcarrier is modulated by $\mathbf{\mathcal{S}}_{A}$ or $\mathbf{\mathcal{S}}_{B}$. For instance, if $\delta_{i}$ is larger, it is more probably that the $i$-th subcarrier is modulated by $\mathbf{\mathcal{S}}_{A}$, whereas if $\delta_{i}$ is smaller, it is more probably to be modulated by $\mathbf{\mathcal{S}}_{B}$. The LLR detector provides near-optimal performance by making hard decisions, while its complexity is significantly lower than the ML detector. 

\section{Proposal of TransD3D-IM Detector\label{sec:DeepNet}}

\vspace{-0.1cm}
This section first describes the operation of the TransD3D-IM detector and provides a comprehensive overview of its structure. Lastly, the offline training procedure is presented for our proposed detector.

\vspace{-0.2cm}
\subsection{Outline of TransD3D-IM\label{subsec:Outline of TransD3D-IM}}

\begin{figure}[tb]
\begin{centering}
\includegraphics[width=\figWidth]{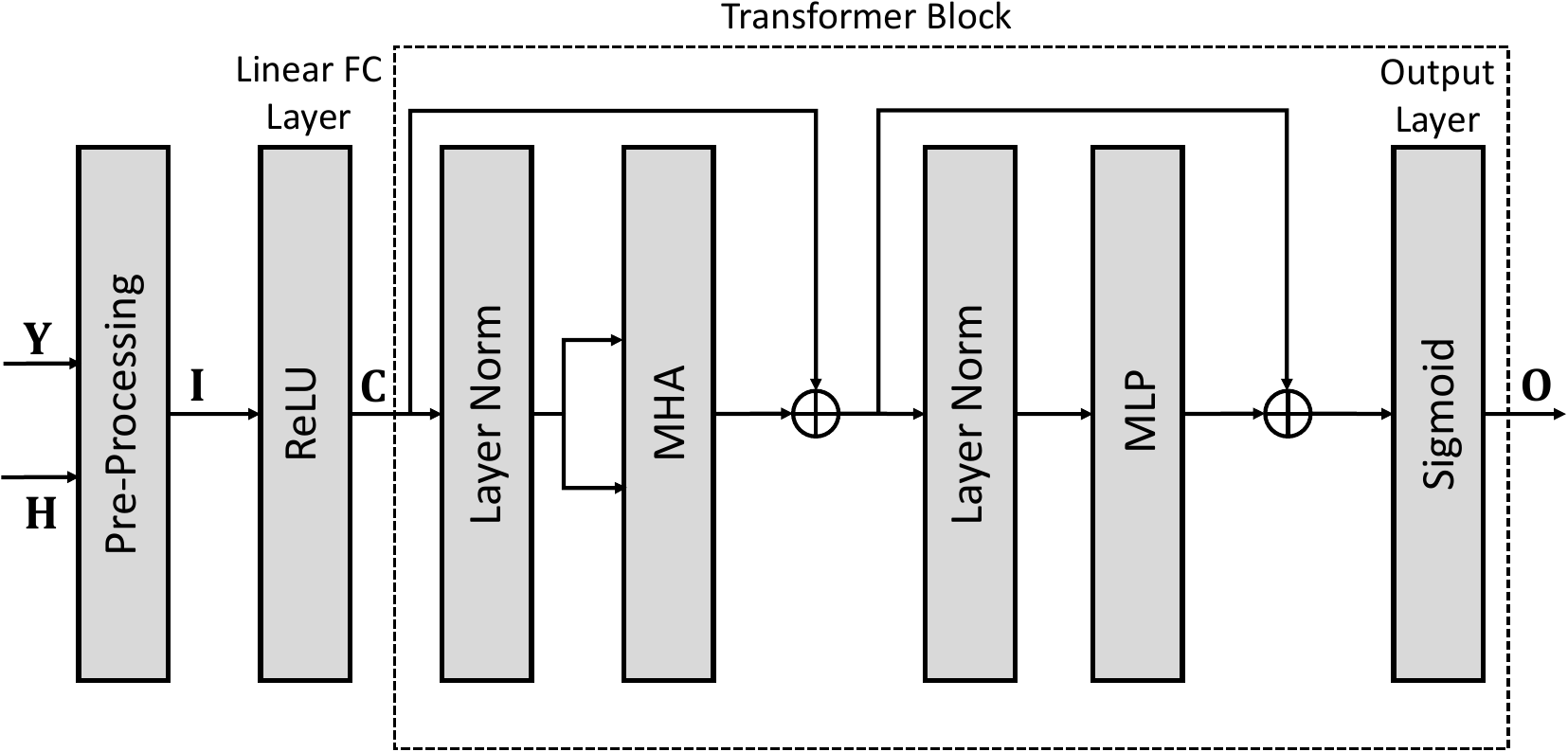} 
\par\end{centering}
\caption{Diagram of the proposed TransD3D-IM detector.\label{fig:figure2}}
\end{figure}

\vspace{-0.1cm}
Firstly, we present a description of the TransD3D-IM detection process. As depicted in Fig.~\ref{fig:figure2}, the received signal and the channel $\mathbf{H}$ are fed to a pre-processing module. The resulting matrix from this module passed through a linear fully connected (FC) layer to produce a matrix with a higher row dimension. Following the linear FC layer, a Transformer block is employed, comprising a multi-head attention (MHA) layer, a multi-layer perceptron (MLP), and layer normalization operations. By performing a sequence of computations and projections within aforementioned DNN blocks, the TransD3D-IM input is transformed into the matrix of identical dimensions, which effectively represents the extracted dependencies and features from the input data. The neural networks have the output data, represented as intermediate matrix $\mathbf{O}$, providing soft probabilities for various symbols before next manipulations. Based on the intermediate matrix $\mathbf{O}$, the indices set of two mapper constellations are selected by the last column of this matrix. Then, the symbols corresponding to their mapped 3D constellation are recovered based on both matrix $\mathbf{O}$ and the previously recovered indices. By combining the process index and symbol parts, the transmitted data could be accurately recovered.

\subsection{Structure of TransD3D-IM\label{subsec:Structure of TransD3D-IM}}
Next, we present the core details of each DNN module in the proposed TransD3D-IM detector.

\textit{Pre-Processing:} Pre-Processing module takes the received signal $\mathbf{Y}$ and the channel matrix $\mathbf{H}$ as inputs and implements zero-forcing (ZF) to build-up an equalized signal vector $\Bar{\mathbf{Y}}$ at the receiver, which is presented by
\begin{equation}
\
\Bar{\mathbf{Y}}=\mathbf{Y}\mathbf{H}^{-1}
. \label{eq:post-process}
\end{equation}

The energy of the received signal, i.e. $\mathbf{Y}_E = [\left|\mathbf{Y}_{1}\right|^{2},\left|\mathbf{Y}_{2}\right|^{2},\left|\mathbf{Y}_{3}\right|^{2}] \in \mathbb{R}^{n\times 3} $, $\mathbf{Y}_{i,\alpha} \in \mathbb{R}^{n \times 1}(1\leq i \leq 3, 1\leq \alpha \leq n)$, is computed to realize various subcarriers clearly. We then concatenate $\Bar{\mathbf{Y}}_R$, $\Bar{\mathbf{Y}}_I$ and $\mathbf{Y}_E$ in parallel and get a matrix $\mathbf{I}=[\Bar{\mathbf{Y}}_R,\Bar{\mathbf{Y}}_I,\mathbf{Y}_E]$, $\mathbf{I}\in\mathbb{R}^{n\times9}$ as the input data of the Linear FC layer, where $\Bar{\mathbf{Y}}_R$ and $\Bar{\mathbf{Y}}_I$ represent the real and imaginary components of the vector $\Bar{\mathbf{Y}}$, respectively.

\textit{Linear Fc layer:} Next, the matrix $\mathbf{I}$ enters a linear FC layer, where employing the rectifier linear unit (ReLU), $f_{\text{ReLU}}\left(x\right)=\text{max}\left(0,x\right)$ as the activation function. This module is implemented to get a matrix $\mathbf{C}$ with a higher row dimension. The resulting matrix $\mathbf{C}$ can be expressed as follows.
\begin{equation}
\
\mathbf{C}=f_{\text{ReLU}}(\mathbf{I}\times\mathbf{W}_{\text{C}}+\mathbf{b}_{\text{C}})
, \label{eq:LinearFC}
\end{equation}
where $\mathbf{W}_{\text{C}}$ and $\mathbf{b}_{\text{C}}$ represent weights and biases to be learned, respectively.

\textit{Transformer Block:} Transformer Block plays a crucial role in our proposed detector. This block is employed to transform vector $\mathbf{C}$ into another matrix, effectively representing the extracted dependencies and features of the transmitted data bits. The details descriptions for every module are described along the following lines.

\begin{figure}[tb]
\begin{centering}
\includegraphics[width=\figWidth]{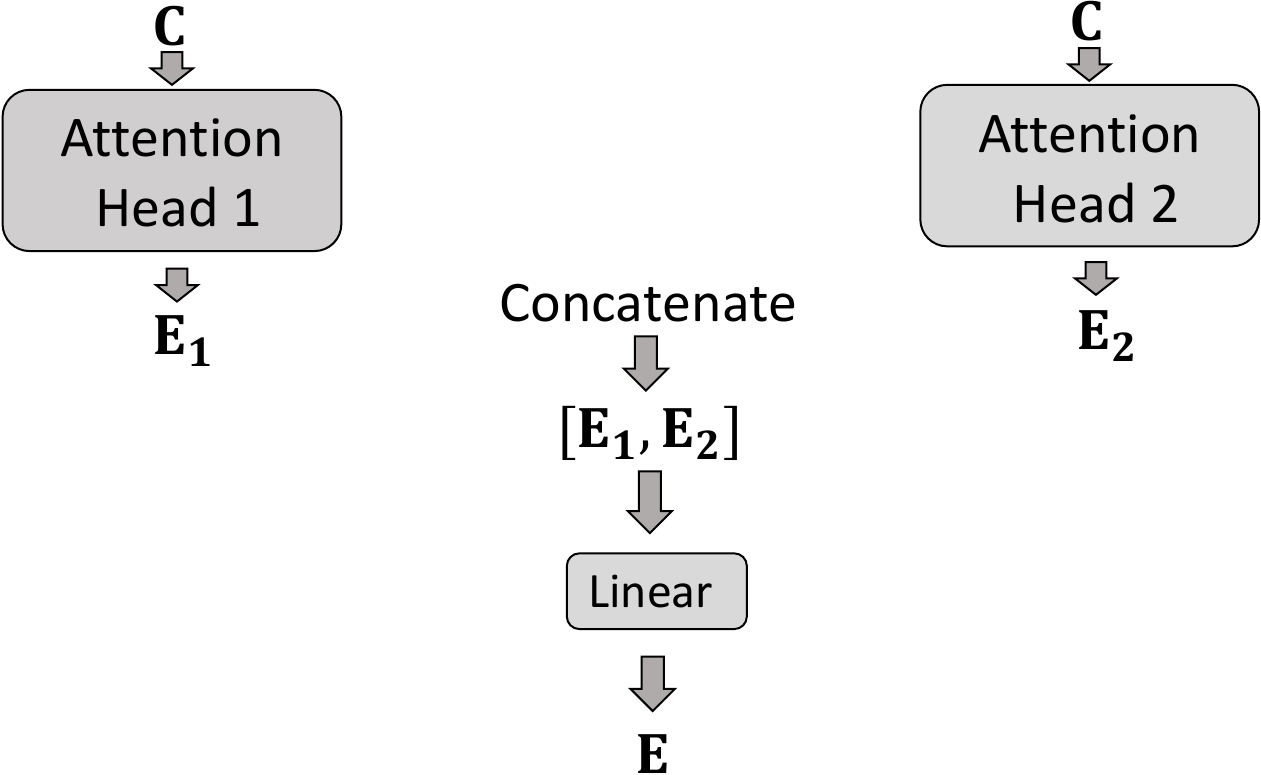} 
\par\end{centering}
\caption{Diagram of two-head attention.\label{fig:figure3}}
\end{figure}

$\bullet$ \textit{Multi-Head Attention Layer (MHAL):} MHA is derived by the single-head self-attention mechanism, which involves the calculation for queries of $\mathbf{Q}$, keys of $\mathbf{K}$, values of $\mathbf{V}$ \cite{attention} and the matrix $\mathbf{E}$. These components are computed as follows. 
\begin{equation}
\begin{cases}
\mathbf{Q}=\Bar{\mathbf{C}}\times\mathbf{W}_{\text{Q}}+\mathbf{b}_{\text{Q}},\\
\mathbf{K}=\Bar{\mathbf{C}}\times\mathbf{W}_{\text{K}}+\mathbf{b}_{\text{K}},\\
\mathbf{V}=\Bar{\mathbf{C}}\times\mathbf{W}_{\text{V}}+\mathbf{b}_{\text{V}},
\end{cases}\label{eq:ZC_sequence}
\end{equation}
Next, the matrix $\mathbf{E}$ of self-attention is computed by
\begin{equation}
\mathbf{E=\text{Softmax}(\mathbf{QK}}^{T})\mathbf{V},
 \end{equation}
where $\mathbf{\Bar{C}}$ is the resulting matrix of $\text{Norm}
\ (\mathbf{C})$ that stands for the function of layer normalization. $\mathbf{W}_\text{Q}$,$\mathbf{W}_\text{K}$,$\mathbf{W}_\text{V}$,$\mathbf{b}_\text{Q}$,$\mathbf{b}_\text{K}$, and $\mathbf{b}_\text{V}$ are weights and biases to be optimized likewise, respectively. $\text{Softmax}\left(\cdot\right)$ is employed as an activation function to implement normalization. In this paper, we employ a two-head attention layer. Therefore, the attention mechanism operated on two project versions of $\mathbf{Q},\mathbf{K},$ and $\mathbf{V}$, producing intermediate attention values of $\mathbf{E}_1$ and $\mathbf{E}_2$, as described in Fig.~\ref{fig:figure3}. The final attention value is obtained by cascading and projecting them once again, which is expressed by
\begin{equation}
\mathbf{E}=[\mathbf{E}_1,\mathbf{E}_2]\times\mathbf{W}_\text{E}+\mathbf{b}_\text{E},
 \end{equation}
where $\mathbf{W}_\text{E}$ and $\mathbf{b}_\text{E}$ represent weights and bias to be optimized of projection, respectively. 

$\bullet$ \textit{Multilayer Perceptron (MLP):} This module comprises two FC sublayers, and ReLU is utilized as the activated function for the first FC sublayer. The output of MLP can be calculated~by
\begin{equation}
\mathbf{A}=f_{\text{ReLU}}(\mathbf{\text{W}}_{\text{A}_{1}}\times\mathbf{\Bar{E}}+\mathbf{\text{W}}_{\text{b}_{1}})\times\mathbf{\text{W}}_{\text{A}_{2}}+\mathbf{\text{b}}_{\text{A}_{2}},
 \end{equation}
where $\mathbf{\Bar{E}}$ is the resulting matrix of $\text{Norm}
\ (\mathbf{E})$, $\mathbf{\text{W}}_{\text{A}_{1}}$, $\mathbf{\text{W}}_{\text{A}_{2}}$, $\mathbf{\text{b}}_{\text{A}_{1}}$, and $\mathbf{\text{b}}_{\text{A}_{2}}$ are params of the first and second FC sublayers, respectively.

$\bullet$ \textit{Layer Normalization:} Layer normalization serves the purpose of stabilizing the distribution of data features, leading to faster model convergence as well as reduced training duration.

\textit{Output layer:} In this layer, the Sigmoid function, $f_{\text{Sigmoid}}(x)=\frac{1}{1+e^{-x}}$ is adopted as the activation function. Accordingly, the obtained matrix $\mathbf{O}$ of the output layer can be expressed by
\begin{equation}
\mathbf{O}=f_{\text{Sigmoid}}(\mathbf{W}_\text{O}\times\mathbf{A}+\mathbf{b}_\text{O}),
 \end{equation}
where $\mathbf{W}_\text{O}$ and $\mathbf{b}_\text{O}$ are weights and bias to be learned, respectively. The matrix of $\mathbf{O}=\left\{\mathbf{O}_1,...,\mathbf{O}_{s_{A}+1} \right\}$ has the size of $[n\times(s_{A}+1)]$, where $\mathbf{O}_{i} \in \mathbb{R}^{n}$ with $1 \leq i \leq s_{A}+1$, provides soft probabilities for the corresponding symbols and the indices set of two mapper constellations based on the output layer of TransD3D-IM, which utilize Sigmoid as the active function that maps outputs in the interval (0,1). In fact, the proposed detector relies on the mechanism of the LLR-aided detection to recovered both indices and data separately corresponding to $\mathbf{O}$.


\vspace{-0.2cm}

\subsection{Model Training\label{subsec:Model Training}}

\vspace{-0.1cm}
Implementing into online TransD3D-IM detector is required sufficiently large offline training. In particular, various $p$-bit sequences are randomly generated and converted to the corresponding conveyed bits $\mathbf{X}$ according to the system model in Section~\ref{sec:System-Model}. Next, the processed signals are passed through the Rayleigh fading channel, where it is subject to interference from random AWGN noise. Channel and noise matrices are also randomly created and changed from one-bit sequence to another, according to their established statistical models. At the receiver, the obtained signal and CSI are pre-processed to obtain the training data.
The conveyed signals $\mathbf{X}$ are transformed to the corresponding one-hot labels to enhance the accuracy of symbol recovery at the receiver side. For instance, suppose that sending a DM-IM-3D-OFDM subblock $[\mathbf{\mathcal{S}}_{A}^{1},\mathbf{\mathcal{S}}_{A}^{2},\mathbf{\mathcal{S}}_{B}^{1},\mathbf{\mathcal{S}}_{B}^{2}]$ with via four subcarriers in Table~\ref{tab:table1}, the resulting one-hot label has the following form
\begin{equation}
    \mathbf{\hat{O}}=\left[\begin{array}{ccc}
1 & 0 & 0\\
0 & 1 & 0\\
1 & 0 & 1\\
0 & 1 & 1
\end{array}\right].
\end{equation}

Here, the row and column numbers of $\mathbf{\hat{O}}$ represent the number of subcarriers $n$ and the constellation size $\mathbf{\mathcal{S}}_A$ and $\mathbf{\mathcal{S}}_B$, i.e., $s_A$ and $s_B$, respectively. In particular, the first two columns indicate the order of 3D symbols in their corresponding constellations. 
Meanwhile, the last column shows the index selectors of the two mapper constellations. The index selector $\mathbf{\mathcal{S}}_A$ corresponds to positions with a value of 0, while positions with a value of 1 indicate an index selector of $\mathbf{\mathcal{S}}_B$.  

The TransD3D-IM model performs a classification task to reduce the discrepancy among one-hot labels $\mathbf{\Hat{O}}$  and the resulting matrix $\mathbf{O}$  of neural networks. Therefore, the binary cross-entropy (BCE) function is adopted as the loss function as follows.
 \begin{equation}
\ell(\mathbf{O},\hat{\mathbf{O}};\mathbf{\theta})=-\mathbf{\hat{O}}\log_{2}(\mathbf{O})-(1-\mathbf{\hat{O}})\log_{2}(1-\mathbf{O}),\label{eq:loss}
\end{equation}
where $\theta$ is parameters to be learned of the model. The Adam optimizer \cite{kingma2014adam} is employed and the learning rate is denoted by $\eta$. Aiming at achieving the best result, the training SNR level, presented as $\lambda_{train}$, must be carefully selected for training TransD3D-IM since it directly affects the detection performance.


\section{Simulation results\label{sec:SIMULATION RESULTS}}

\vspace{-0.1cm}

In this section, we provide BER performance comparisons and runtime complexity of TransD3D-IM with its competitors, particularly ML, LLR, and the existing DL-based detector, term as DuaIM-3DNet \cite{DANGY2022}. As for TransD3D-IM, the linear FC layer has $q_{\text{linear FC}}$ nodes. We consider a two-head attention layer and implement MLP with two FC layers, which have $q_{{\text{MLP}}}$ nodes. Meanwhile, the quantity of the hidden nodes of two subnets in the DuaIM-3DNet are denoted by $q_\text{IndexNet}$ and $q_\text{SymbolNet}$, respectively. To ensure that DL-based detectors have similar complexity, $q_\text{IndexNet}$ and $q_\text{SymbolNet}$ have been used parameters as in \cite{DANGY2022}. These DL-aided detectors are trained for 100 epochs with every epoch consisting of 600 batches. The batch size used during training is set to 1000 epochs. The learning rate of $\eta=0.001$ is applied for all parameters. In this paper, we consider the case with $(n,k)=4,2$ and use two types of params for all scenarios, namely Scenario 1 and 2, as shown in Table~\ref{tab:table2}. Our analyses are considered in scenarios with perfect CSI conducted in the Rayleigh fading channel where each entry is distributed by the complex Gaussian $\mathcal{C}\mathcal{N}(0, 1)$.

\begin{table}[!ht]
\centering
\caption{Parameters For DNN-Based Detectors\label{tab:table2}}
\begin{tabular}{|c|c|c|c|c|c|c|}
\hline 
 & $s_{A}$ & $s_{B}$ & \multicolumn{2}{c|}{DuaIM-3DNet} & \multicolumn{2}{c|}{TransD3D-IM} \\
\cline{4-7}
 &  &  & $q_{\text{IndexNet}}$ & $q_{\text{SymbolNet}}$ & $q_{\text{Linear FC}}$ & $q_{\text{MLP}}$ \\
\hline 
Scenario 1 & 2 & 2 & 768 & 768 & 32 & 128 \\
\hline 
Scenario 2 & 4 & 4 & 1000 & 1000 & 64 & 256 \\
\hline 
\end{tabular}
\end{table}

\label{subsec:berPer}

\vspace{-0.1cm}

\subsection{BER Performance}

\label{subsec:berPer} 
\begin{figure}[!ht]
    \centering
    \includegraphics[width=\figWidth]{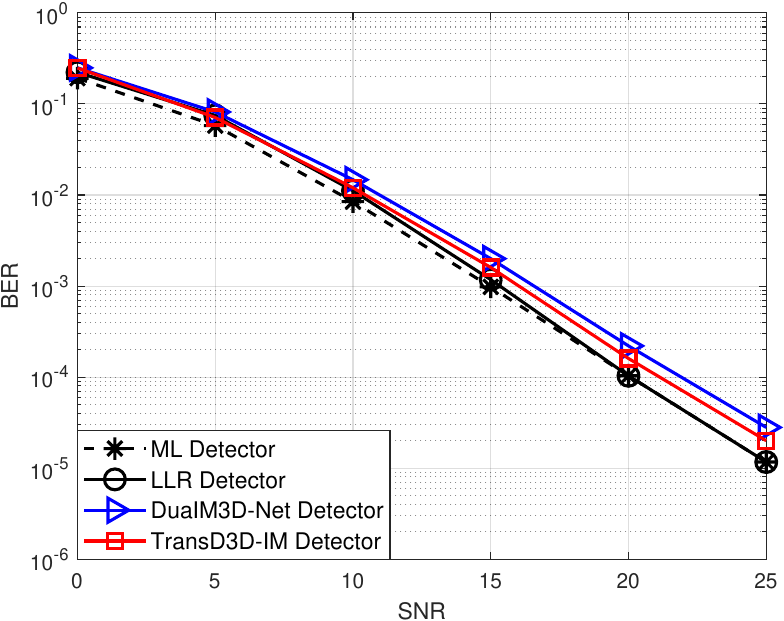}
    \caption[]{BER performance of our proposed TransD3D-IM detector and the existing ones for Scenario 1.}
    \label{fig4}
\end{figure}

In Fig.~\ref{fig4},
we evaluate BER of TransD3D-IM in comparison with the current detectors versus SNR for Scenario 1 under
perfect CSI channel condition. The DL-based detectors have been trained with $\lambda_{train}=15$ dB. It can be seen that the proposed TransD3D-IM detector attains the BER performance approaching the model-based detectors and is slightly superior to the DuaIM-3DNet. For instance, the gap SNR between the Transformer-aided DL detector, LLR, and ML is negligible with less than 1 dB and 2 dB at BER of $10^{-3}$, respectively.

\begin{figure}[!ht]
    \centering
    \includegraphics[width=\figWidth]{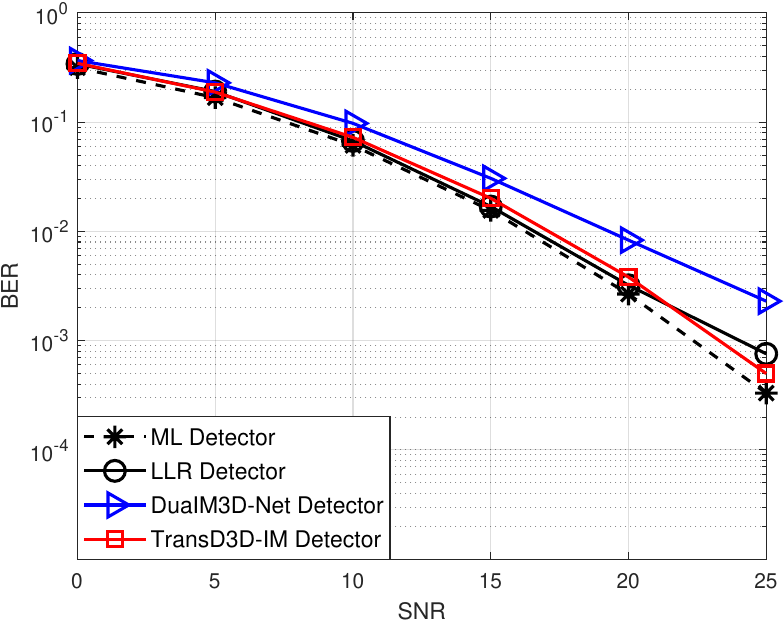}
    \caption[]{BER performance of the proposed TransD3D-IM and existing detectors for Scenario 2.}
    \label{fig5}
\end{figure}

Fig.~\ref{fig5} shows the BER performance of TransD3D-IM comparisons to its competitors versus SNR for Scenario 2 under certain CSI conditions. As shown in Fig.~\ref{fig5}, we observed that TransD3D-IM achieves BER performance approximate to that of the model-aided detectors while significantly outperforming the existing DL-based detector. Considering at BER of $10^{-2}$, the proposed detector is only inferior to ML and LLR by about 1dB. Meanwhile, it has an edge over DuaIM-3DNet by about 5~dB. This indicates that differences in performance between various DL-aided detectors are increasingly evident. Our simulation results point out that TransD3D-IM can achieve nearly optimal BER performance in comparison with model-based receivers. Furthermore, our proposal exhibits more robustness than the current DL-aided detector while TransD3D-IM has significantly lower runtime complexity compared to its equivalents as detailed in the next section.


\vspace{-0.1cm}

\vspace{-0.1cm}

\subsection{Runtime Complexity}

\label{subsec:ComComparision} We now compute the runtime complexity required for signal detection of the proposed detector and its competitors. All candidates are simulated using MATLAB on the same computer. The system and DNN model parameters are employed as presented in the previous subsection. We show the comparison of the resulting runtime of all models in Table~\ref{tab:comlex}. It is clear that TransD3D-IM requires much less runtime than its counterpart under the same condition. Specifically, the runtime of TransD3D-IM is 460 times less than that of ML and 5 times less than that of LLR and DuaIM-3DNet detectors. This unequivocally validates the advantage of our proposal in terms of runtime.

\begin{table}[!ht]
\centering
\caption{Runtime of proposed TransD3D-IM and its competitors in milliseconds\label{tab:comlex}}
\begin{tabular}{|c|c|c|c|c|}
\hline 
$(n,k,s_{A},s_{B})$ & ML & LLR & DuaIM-3DNet & TransD3D-IM\tabularnewline
\hline 
\hline 
(4,2,4,4) &$12$  &$0.134$  &$0.136$  &$0.026$ \tabularnewline
\hline 
\end{tabular}$ $
\end{table}

 
\section{Conclusions\label{sec:Conclusions}}

\vspace{-0.1cm}
In this work, we have put forward the TransD3D-IM detector for the DM-IM-3D-OFDM system. Specifically, we build a novel DNN model for TransD3D-IM based on the Transformer framework. Our proposal can comprehensively leverage the global dependencies and features of the obtained signal to produce nearly optimal BER performance. Under the Rayleigh fading channel with CSI certainty, TransD3D-IM can attain approximate BER performance in comparison with the model-based receiver. Besides, the proposed detector exhibits more robustness than the existing DL-aided method, especially when the model system parameters increase. Meanwhile, TransD3D-IM requires lower runtime complexity than its competitors. We believe that with such advantages, Transformer-based detectors may be an effective solution for signal detection in communication systems in the future.
	 \bibliographystyle{IEEEtran}
 \balance
\bibliography{IEEEabrv,refs}

\end{document}